# Development of Active Pixel Photodiode Sensors for Gamma Camera Application

*Nur SULTAN SALAHUDDIN* [1,2], *Michel PAINDAVOINE* [1], *Brahmantyo HERUSETO* [2], *Michel PARMENTIER* [3]

[1]LE2I Laboratory UMR CNRS 5158, University of Burgundy, Dijon, France.
[2]Gunadarma University, Jakarta, Indonesia.
[3]Laboratoire Imagerie et Ingénierie pour la santé, University of Franche-Comte, Besançon, France.

We designed new photodiodes sensors including current mirror amplifiers. These photodiodes have been fabricated using a CMOS 0.6 micrometers process from Austria Micro System (AMS). The Photodiode areas are respectiveley 1mm x 1mm and 0.4mm x 0.4mm with fill factor 98 % and total chip area is 2 square millimetres. The sensor pixels show a logarithmic response in illumination and are capable of detecting very low blue light (less than 0.5 lux) . These results allow to use our sensor in new Gamma Camera solid-state concept.

## I. Introduction

In recent years[1] there has been a growing interest in developing compact gamma cameras to improve nuclear medicine imaging. Conventional full-size gamma cameras using a NaI (TI) scintillator block coupled to a bulky array of PMT are, by nature of their large size, preclude from use in more clinic situation. There are three major design approaches to the development of compact gamma cameras: (1) discrete scintillator/photodiode cameras wherein the gamma-rays interact in an array of optically isolated scintillation crystals coupled 1-to-1 to an array of solid-state photodiode[2]-[5]; (2) solid-state cameras where the gamma-rays interact directly with a pixellated solid-state detector such as CdZnTe [6]-[8]; and (3) position-sensitive photo multiplier tube (PSPMT) cameras where the gamma-rays interact in one or more scintillation crystal which are subsequently read out by a single PSPMT [9]-[14]. The compact scintillation camera uses an array of discrete scintillator crystals and a matching array of photodiodes to detect the scintillation light that result when a gamma-ray is absorbed [3]. This scheme thus replaces the bulky PMT photodectectors used in conventional scintillation cameras with small photodiodes, greatly reducing the camera size as shown in figure 1 In addition, the scintillator CsI (TI) can be used with photodiodes (but not with PMTs) and replaces the NaI (TI) used in conventional cameras, allowing another decreases in camera size since NaI (TI) requires special, bulky packing but CsI (TI) does not.

The compact scintillation/photodiode cameras offer several advantages over conventional scintillation cameras: (1) array of small photodiodes provide improved intrinsic spatial resolution; (2) the small camera size allows shorter imaging distances, thus improving collimator resolution; (3) the compact design permits a greater variety of viewing angles and allows multiple cameras to take different views simultaneously; and (4) the multiple scintillator photodiode channels yield a higher overall maximum event rate; (5) the smaller size lowers cost. The two advances that now make discrete scintillation camera technology a viable option for scintimammography applications are the low leakage current (~50 pA/pixel) photodiode array and the custom IC readout of the photodiode signals. This is important to achieving a compact, cost effective design, because with the many pixels that will be present in a complete camera, discrete electronics become prohibitively bulky and expensive.

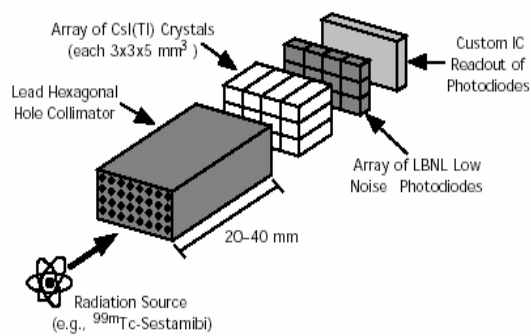

Figure 1. Module of discrete scintillation camera.

In this context we designed a new CMOS image sensor array that we present in this article. We introduce in the second section scintillator photodiode detectors theory. In the third and fourth sections we describe our design of our CMOS Active Pixel Sensor dedicated to Gamma Camera Imaging. In the fifth section we present fabrication and test results about this new sensor.

## II. Pixel Design Description

In a standard CMOS process several parasitic junction can be used as photodiode aither p-well or n-well[15]. two pixels different structure as shown in figure 2 have been realized using a photodiode formed n+ difusion in the n-well (vertical photodiode), n+ and p+ diffusion in the n-well ( lateral photodiode) and vertical-lateral photodiode combination. The photodiode is forward biased, and when incoming photons are absorbed, a photocurrent proportional to the intensity of light flows through the photodiode. This current is converted to output voltage value using current mirroring integration readout circuits[16].

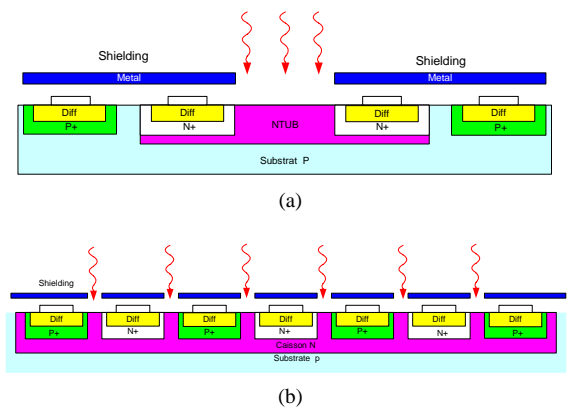

Figure 2. Two Photodiode structures on p-type substrate, (a) partly hollowed n+ diffusion in n-well (vertical photodiode), (b) partly hollowed n+ and p+diffusion in n-well (lateral photodiode.

## III. Fabrication and Test Results

Figure 3 shows a layout and photograph of small test pixels implemented in a standard 0.6 μm CMOS process from AMS. The chip consists mainly of current-mirror amplifiers circuit and active pixel sensors (APS) with different sizes and structures : (a) 1x1 mm$^2$ (the three pixels: vertical, lateral and vertical-lateral combination ), (b) 400x400 μm$^2$ (the four pixels:vertical).

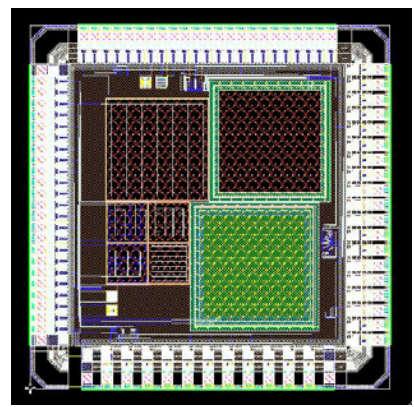

(a)

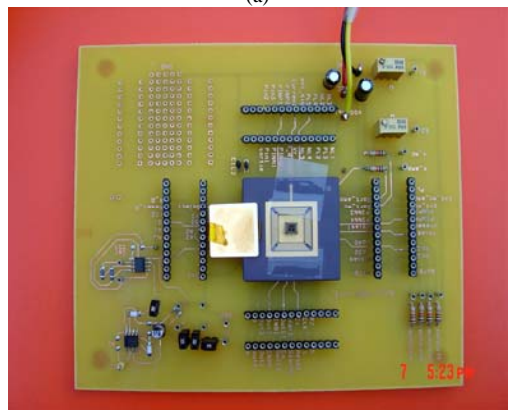

(b)

Figure 3. (a) Layout Of the Detector Pixel with a 2mm x 2mm area in a 0.6 μm CMOS process, (b) photograph of photodiode sensors.

The photo response of test photodiodes on the chip is obtained by measuring the photo current under illumination from 400 nm to 720 nm.

Figure 4a shows responsibility of the photodiodes sensors to a monochromatic. Figure 4b shows the response of the photodiodes sensors under illumination in the blue light (400 nm).

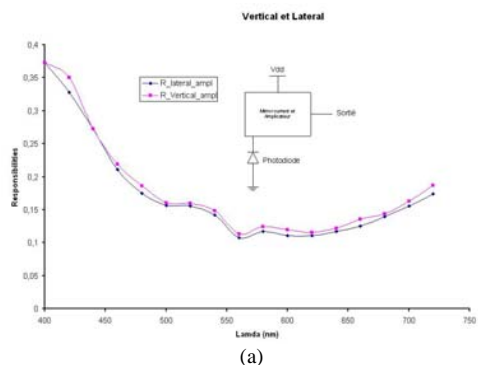

(a)

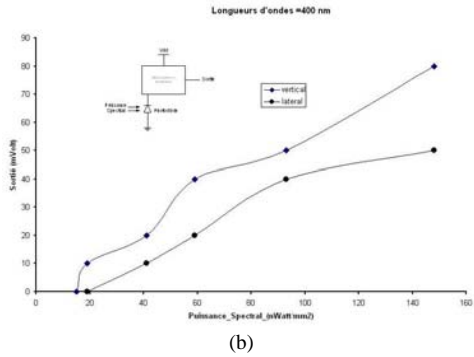

(b)

Figure 4. The response the photodiodes sensors

Figure 4 shows variations of photo current in dynamic mode. In this mode, light emitting diode is driven with a pulse generator. Results are presented in figures 4a and 4b are obtained with a blue LED calibrated for a 0.5 lux. In these figures the upper curves represent the LED voltage input and the lower curves represent the sensor output.

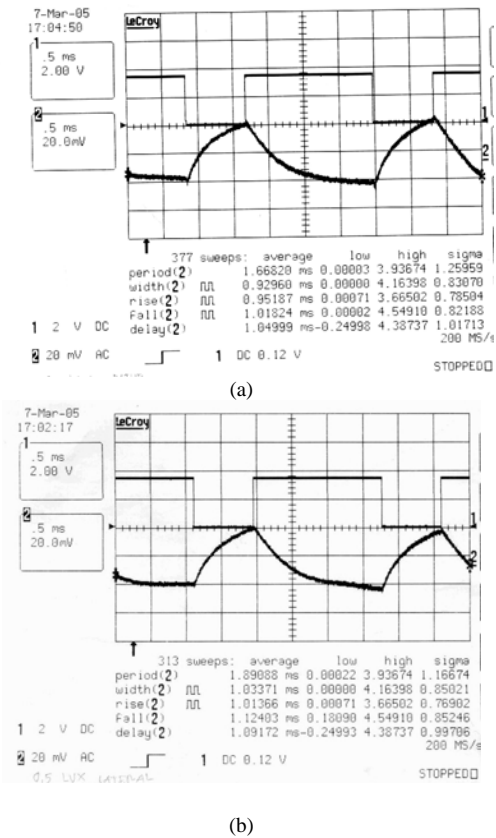

(b)

Figure 4. Photocurrent in dynamic mode :(a) vertical photodiode 1mm x 1mm area and (b) lateral photodiode 1mm x 1mm area

In order to measure the capacity of the photodiodes, we use circuit shown in figure 5. In this mode, voltage source is driven with a pulse generator. Results are presented in figures 6.

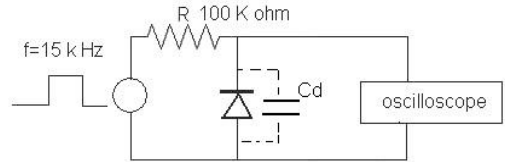

Figure 5. Photodiode used in integrator mode.

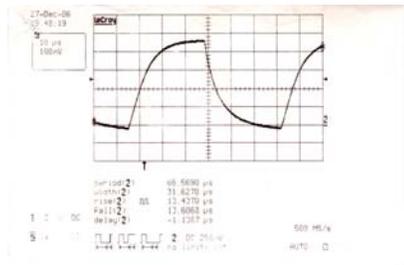

Figure 6. The measurement of the capacitance photodiodes.

The voltage V discharges with the time-constant:
$$t = R \times Cd$$

Obtained the values for the capacitance photodiodes are respectiveley 65 pF and 35 pF for 1mm x 1mm and 0.4mm x 0.4mm. Table 1 summarizes the overall measurement characteristics.

Table 1-. The measurement characteristics

| Parameter | Result |
|---|---|
| Technology | 0.6 µm CMOS, 2-layer Metal et 1-layer Poly |
| Photodectectors | Diffusion N+ in Nwell/ P Substrat (vertical Photodiode), Diffusion N+ and P+ in Nwell (lateral Photodiode) |
| Pixel Pitch | 1 mm x 1mm, |
| Typical Capacity | 65 pF (1 mm x 1 mm), 35 pF (0.4 mm x 0.4mm) |
| Fill Factor | 98 % |
| Spectral Response | 460 nm (BLUE) |
| Power Supply | 5 Volt |
| Response in Illumination | Logarithmic |

## IV. Conclusions and Perspectives

The CMOS active photodiode sensor and current mirror amplifier has been fabricated using a 0.6 µm CMOS process. The experimental results show that this sensor has logaritmic response in illumination and is capable of detecting very low blue lights emitting diode. These results allow us to consider using of this tehnology in new solid state gamma cameras.